\documentstyle[12pt]{article}
\input epsf

\newcommand{\eqref}[1]{Eq.~(\protect\ref{#1})}

\begin{document}

\begin{center}
{\bf Elasticity Theory of a Twisted Stack of Plates}
\end{center}

\begin{center}
C. S. O'Hern\footnote{ohern@lubensky.physics.upenn.edu}, Randall D.
Kamien\footnote{kamien@dept.physics.upenn.edu}, T. C.
Lubensky\footnote{tom@lubensky.physics.upenn.edu}, and Philip
Nelson\footnote{nelson@dept.physics.upenn.edu} \\
Department of Physics and Astronomy,
University of Pennsylvania \\
Philadelphia, PA~~19104\\
July 2, 1997
\end{center}

\bigskip

\begin{center}
{\bf Abstract}
\end{center}

We present an elastic model of B-form DNA as a stack of thin, rigid
plates or base pairs that are not permitted to deform.  The symmetry
of DNA and the constraint of plate rigidity limit the number of bulk
elastic constants contributing to a macroscopic elasticity theory of
DNA to four.  We derive an effective twist-stretch energy in terms of
the macroscopic stretch $\epsilon$ along and relative excess twist
$\sigma$ about the DNA molecular axis.  In addition to the bulk
stretch and twist moduli found previously, we obtain a twist-stretch
modulus with the following remarkable properties: 1) it vanishes when
the radius of the helical curve following the geometric center of each
plate is zero, 2) it vanishes with the elastic constant $K_{23}$ that
couples compression normal to the plates to a shear strain, if the plates
are perpendicular to the molecular axis, and 3) it is nonzero if the
plates are tilted relative to the molecular axis.  This implies that a
laminated helical structure {\it carved} out of an isotropic elastic
medium will not twist in response to a stretching force, but an
isotropic material will twist if it is {\it bent} into the shape of a
helix.

\smallskip

\noindent {\sl PACS:
87.15.-v, %  Molecular biophysics
87.10.+e, %  General, theoretical, and mathematical biophysics
87.15.By.%  Structure, bonding, conformation, configuration, and isomerism of
%%biomolecules
}

\newpage

\section{Introduction}

The elastic properties of DNA have become a focus of recent research
\cite{Siggia,stretchdna,fain,Markonew,Kamien,Mezard}.  In
particular, stretching experiments on single molecules of DNA now
provide a direct probe of the bending, stretching, and twisting
elasticity of DNA\cite{smith2,Wang,Strick}.  Understanding the
elasticity of single molecules of DNA may be relevant {\it in vivo}; {\it
e.g.}, the recA protein which is responsible for homologous
recombination in bacterial mitosis has been observed to stretch and
twist DNA when it is bound to the DNA molecule\cite{stasiak}.

Marko and Siggia recently modeled DNA as a thin, uniform rod with a
linear bending elasticity and calculated the extension of the rod as a
function of the applied stretching force\cite{stretchdna}.  This
worm-like chain model was appropriate for stretching forces $f \ll
\gamma$, where $\gamma \sim A k_B T/R^2 \approx 200{\rm pN}$ is the
stretch modulus, $A \approx 50~ {\rm nm}$ is the bend persistence
length, and $R = 1~{\rm nm}$ is the radius of the molecule.  Forces in
this regime pull on the thermal fluctuations of the molecular backbone
but do not pull on the internal structure of DNA.  The worm-like chain
model accurately predicts the extension of the molecule to within
$10$\% up to approximately $10$~pN\cite{smith2}.  Above
$10$~pN\cite{smith2,Wang} there are no longer thermal effects and DNA
is stretched {\sl elastically}.  A strain variable $\epsilon$ must be
introduced to describe increases in the molecular length after
thermally induced contour length fluctuations are fully
stretched\cite{stretchdna,cluzel,smith2,Wang}.

DNA is also characterized by the degree of twist of its phosphate
backbones about the central axis.  Relative twist excess or deficit
from a state of preferred twist is parameterized by a signed twist
variable $\sigma$.  The chiral asymmetry of DNA allows for a
twist-stretch coupling between $\sigma$ and $\epsilon$; this coupling
is clearly visible in the recent experiments of Strick, {\it
et. al.}\cite{Strick,Markonew,Kamien}.  Theoretical work has focused
on calculating the extension of the molecule as function of both the
applied stretching force and the applied excess
twist\cite{fain,Mezard,Moroz}.  However, in this paper
we investigate a more microscopic origin of the twist-stretch
coupling\cite{philnote}.

We introduce a simple elastic model of B-form DNA in which the
molecule is viewed as a stack of thin, rigid plates that represent
base pairs and are rotated and displaced relative to each other.  (See
Section $3$ below for a review of the geometry of B-form DNA.)  The
centers of mass of the plates define a helical path around a straight
central line, which we call the molecular axis.  The perpendicular
distance from the molecular axis to the helix is called the helix axis
offset.  Distortion energies in this model are controlled by an
underlying continuum elastic energy with those elastic constants
allowed by the anisotropic symmetry of DNA.  Our model of DNA as a
collection of thin, rigid plates has four contributing elastic
constants coupling strains with spatial variation normal to the
plates.  Of particular importance is the elastic constant $K_{23}$
coupling compression perpendicular to the plates to a shear
displacement parallel to the plates.  $K_{23}$ vanishes if the
phosphate backbones point in the same direction\footnote{Each phosphate
backbone has a particular orientation: either $3$'-$5$' or $5$'-$3$'.}
and also vanishes in the isotropic limit in which each plate has
reflection and ${\rm C}_4$ or higher symmetries.

We calculate the twist-stretch coupling in terms of its elastic
constants and the small helix axis offset.  We find that the
twist-stretch coupling has terms linear and quadratic in the helix
axis offset.  The coefficient of the linear term is proportional to
the elastic constant $K_{23}$.  In the isotropic limit or when
$K_{23}$ vanishes, the twist-stretch coupling is quadratic in the
helix axis offset as we calculated earlier\cite{Kamien,Ohern} in
related but simpler models.

An important simplification of our model is the constraint that the
plates comprising the DNA stack are rigid and undeformable.  This is
equivalent to setting to infinity all elastic constants coupling the
strains within a plate.  We thus neglect propeller, buckle, opening,
{\it etc}.  deformations\cite{Lavery} of individual base pairs and
focus instead on inter-base pair deformations described by relative
rotations (roll, tilt, and twist) and translations (slide, shift, and
rise) of the base pairs\cite{Calladine,Lavery}.  We believe this is a
reasonable approximation for stretching forces in the range $1~{\rm
pN} < f < 10$~pN and relative twist excesses $\sigma < 0.05$.

\section{Description of the Model}

\begin{figure}
\epsfxsize=4.5truein
\centerline{\epsfbox{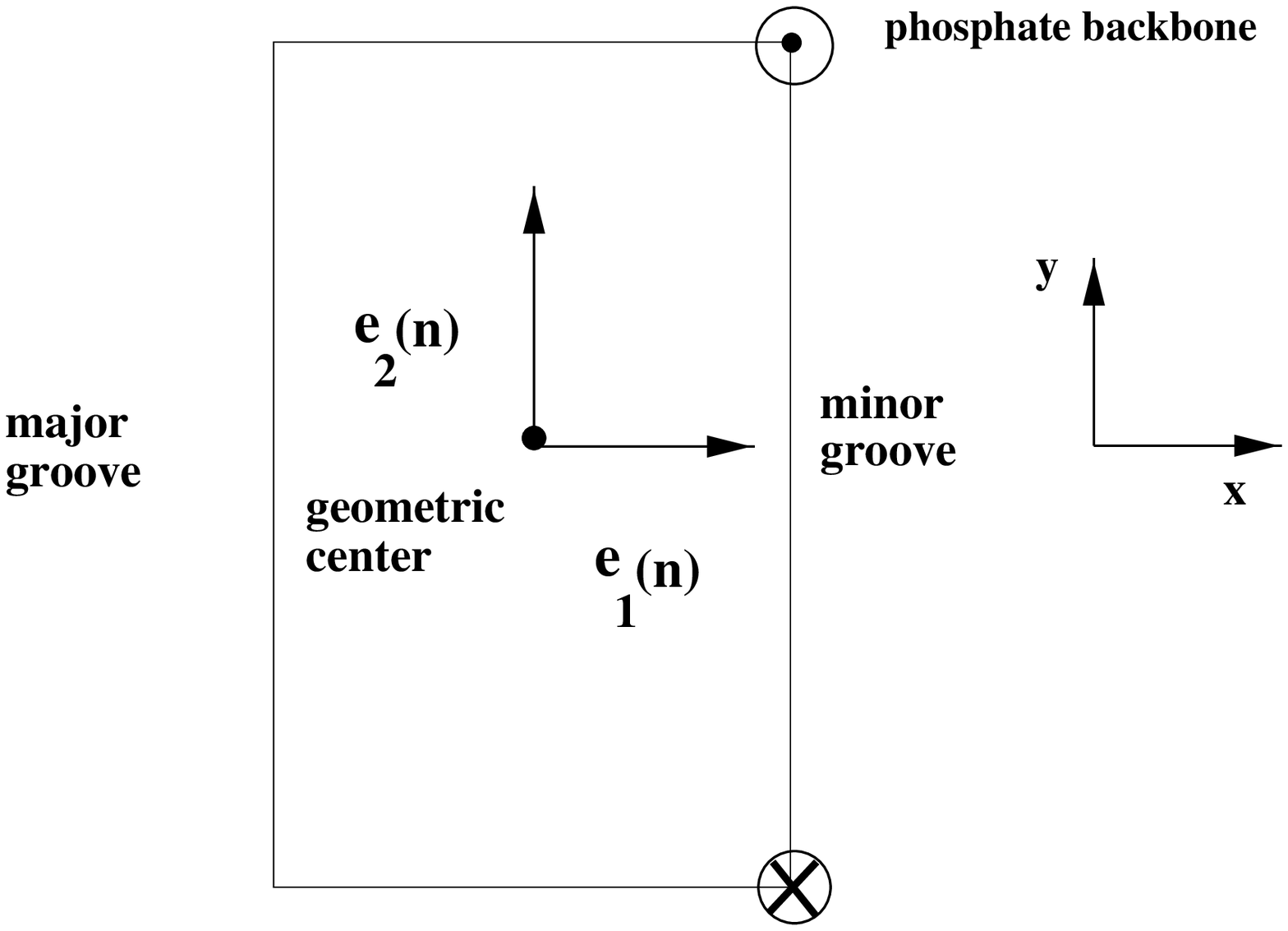}}
\caption{
Schematic cross section of DNA taken perpendicular to the sugar-phosphate
backbones.  ${\bf e}_1$ is perpendicular to the line connecting the two
phosphate backbones and points from the geometric center of the plate
toward the minor groove and ${\bf e}_2$ points from the geometric
center toward the backbone running in the positive ${\bf e}_3$
direction.  Each slice has a coordinate system $(x, y, z)$ with ${\bf
e}_1$ as the $x$-axis, ${\bf e}_2$ as the $y$-axis, and ${\bf e}_3$ as
the $z$-axis.  Also, each slice is labeled by an integer $n$.
}
\label{fig1}
\end{figure}

DNA is composed of base pairs connected by two oppositely-directed
sugar-phosphate backbones that wrap in two helices around the central
molecular axis to produce the major and minor grooves.  A
cross-section of DNA normal to the phosphate backbones is shown in
Fig.\ref{fig1}.  To construct our model for DNA, we first imagine
unwinding the helix so that the phosphate backbones describe two
straight, parallel paths.  This unwound structure
is an elastic rod characterized by an elastic constant tensor
$K_{ijkl}$ with components constrained by the symmetry of the rod.
Each cross-section of the rod looks like Fig. \ref{fig1}.  Let the
$z$-axis be parallel to the phosphate backbones, and let the $x$-axis
pass through the midpoint and perpendicular to the line connecting the
backbones. The only symmetry operation of this rod is a rotation
through $\pi$ about the $x$-axis that causes the axes to transform as
$x \rightarrow x$, $y
\rightarrow - y$, and $z \rightarrow -z$.  This symmetry allows for a
maximum of $13$ independent elastic constants.  Our assumption of the
rigid shape of each cross-section in the $x-y$ plane eliminates from
consideration all strains that require spatial derivatives with
respect to $x$ or $y$.  Thus we need only consider the strains
$u_{zz}$, $u_{zx}$, and $u_{zy}$ and the four elastic constants
$K_{33} = K_{zzzz}$, $K_{11}= K_{zxzx}/4$, $K_{22} = K_{zyzy}/4$, and
$K_{23} = K_{zyzz}/2$ permitted by symmetry that couple these strains.
The elastic constant $K_{23}$ is allowed because the two phosphate
backbones are oppositely-directed.  If they were undirected or pointed
in the same direction, the reflection $y\rightarrow -y$ would be a
symmetry and $K_{23}$ would be zero.  If, in addition, a rotation by
$\pi/2$, which transforms $x\rightarrow y$ and $y\rightarrow -x$, were
a symmetry, there would be only two elastic constants $K_{33}$ and
$K=K_{11}=K_{22}$.  We will refer to this case with two elastic
constants as the isotropic limit.  We view the cross-sectional slices
as rigid plates that are coupled elastically via the elastic constants
$K_{33}$, $K_{11}$, $K_{22}$, and $K_{23}$.

We now imagine that chiral forces inherent to DNA distort the straight
rod to a helical structure in which the centers of mass of the rigid
plates describe a helical path about a straight helical axis.  This
helical structure can be produced by displacing and rotating
neighboring plates by a constant amount.  We assume that chiral
energies leading to a ground-state helical structure are linear in
strain ({\it i.e} surface terms) so that the energies of distortions
from the ground state are determined by the elasticity of the original
untwisted rod, {\it i.e.}, by the elastic constants $K_{33}$,
$K_{11}$, $K_{22}$, and $K_{23}$.  In the following sections, we will
derive the twist-stretch coupling in terms of these constants and the
helix axis offset.

\section{Geometry of B-Form DNA}

We now consider three different ways of stacking our thin, rigid
plates to create a helical model of the straight state of
B-DNA.  We first imagine simply twisting the plates about the long
axis passing through the center of mass of the original rod to create
shape I.  In this shape, the geometric centers of each slice lie on
top of one another, the slices are rotated about the long axis
relative to one another, and the slices remain perpendicular to the
long axis.  Shape II is obtained by first twisting the plates about
the long axis and then removing material from the side of each plate
opposite to the phosphate backbones.  The line connecting the new
geometric centers of each slice follows a helix.  The slices remain
perpendicular to the helix axis about which the geometric centers of
each plate rotate.  (In the discussion below we use the terms helix
axis and DNA molecular axis interchangeably.)  Shape III is obtained
by bending the long axis of the sliced rod into a helix.  The curve
${\vec r}$ connecting the geometric centers of each plate again
follows a helix, but now the plates are perpendicular to the local
tangent to ${\vec r}$ instead of the helix axis.

We note that in the unstressed state of B-form DNA the plates are
roughly perpendicular to the molecular axis, and their geometric
centers nearly coincide with the molecular axis\cite{saenger}.  Thus
to study small deformations of B-DNA, it suffices to restrict our
attention to the case where shapes II and III are small perturbations
of shape I.  In shape II, the amount of material removed from each
slice and hence the radius of the helix is small and in shape III the
bend of the rod axis away from the molecular axis is also small.
Shapes II and III can be described as twisted stacks of plates with
small helix axis offsets.  If the collection of plates has a helix
axis offset, the plates are not stacked with one directly on top of
the other; instead, the shift vector has a component perpendicular to
the molecular axis.  We will see that in our model the helix
axis offset is the origin of the twist-stretch coupling.

\section{Elasticity Theory}

In what follows we will develop a long-wavelength elasticity theory
for a collection of stacked, rigid plates in terms of small deviations
of the shifts and rotations away from their unstressed values.  We
will then eliminate the shift and rotation variables in favor of the
stretch along the molecular axis $\epsilon$ and the relative excess
twist about the molecular axis $\sigma$.  This will allow us to find
the bulk twist and stretch moduli in terms of four elastic constants
and geometric properties of the molecule.  More importantly, it will
also enable us to calculate the twist-stretch coupling and determine
how it scales with the small helix axis offset.

As shown in Fig.~\ref{fig1} we inscribe on each plate a right-handed,
orthonormal triad ${{\bf e}_{\alpha}(n)}$ where $\alpha = 1,2,3$ and
$n$ is a unitless parameter which labels the plate.  In the following,
Roman indices run over Cartesian space coordinates $(x,y,z)$ and Greek
indices label the plate-fixed coordinate system ($1$,$2$,$3$).  ${\bf
e}_1$ and ${\bf e}_2$ lie in the plane of the plate, and ${\bf e}_3$
is perpendicular to the plate.  The coordinates on each plate are
labeled by the pair $(\eta_1, \eta_2)$ which corresponds to the point
on the plate $\eta_1 {\bf e}_1(n) + \eta_2 {\bf e}_2(n)$.  We must
also specify the origin on each plate.  We choose the origin to be the
geometric center ({\it i.e.}, center of mass) of the plate, but this
choice is arbitrary and does not affect the elasticity theory, as we
will discuss in the Appendix.  We can now describe the trajectory in
space of the plates and the associated plate-fixed triads.  We allow
two sets of parameters: a vector of rotation rates ${\vec
\omega}$ and shifts ${\vec
\delta}$.  We decompose ${\vec \omega}$ and ${\vec \delta}$ in the plate-fixed
basis ${\bf e}_{\alpha}(n)$ and assume that the components
$\omega_{\alpha}$ and $\delta_{\alpha}$ are constants independent of
$n$.
\begin{equation}
\begin{array}{ll}
{\vec \omega} = \omega_{\alpha} {\bf e}_{\alpha}(n) \\
{\vec \delta} = \delta_{\alpha} {\bf e}_{\alpha}(n),
\end{array}
\hspace{0.2in} {\rm where} \hspace{0.1in} \alpha = 1,2,3.
\label{body}
\end{equation}
The vector of rotation rates ${\vec \omega}$ describes the rate of change of
the orientation of neighboring plates, and thus
\begin{equation}
{d{\bf e}_{\alpha} \over dn} = {\vec \omega} \times {\bf e}_{\alpha} =
-\epsilon_{\alpha \beta \gamma}
\omega_{\beta}
{\bf e}_{\gamma},
\label{triad}
\end{equation}
where $\epsilon_{\alpha \beta \gamma}$ is the antisymmetric tensor.
The second set of parameters ${\vec \delta}$ describes the relative
displacement of two neighboring plates.  We choose ${\vec \delta}$ to
be the relative displacement of the {\it origin} ${\vec r}(n)$ of the
$(\eta_1, \eta_2)$ coordinate system.  We have
\begin{equation}
{d {\vec r}(n) \over dn} = {\vec \delta} = {\bf e}_{\alpha} \delta_{\alpha}.
\label{tangent}
\end{equation}
Note that according to our definitions ${\vec \omega}$ is unitless and
${\vec \delta}$ has units of length.  Finally, the position in space of the
point $(\eta_1, \eta_2)$ on plate $n$ is simply ${\vec x}(\eta_1,
\eta_2, n) = {\vec r}(n) + \eta_1 {\bf e}_1(n) + \eta_2{\bf e}_2(n)$.
For constant $\omega_{\alpha}$ and $\delta_{\alpha}$, these equations
will, in general, describe a helical structure.
We may solve (\ref{triad}) and
(\ref{tangent}) to find:
\begin{equation}
{\vec r}(n) = {\bf e}_{\beta}(0)\Biggl\{ n {\omega_{\beta}
({\vec \delta} \cdot {\vec \omega}) \over \omega^2} +
\Biggl( \delta_{\beta} - \omega_{\beta}
{ ({\vec \delta} \cdot {\vec \omega}) \over \omega^2 } \Biggr)
{\sin(|{\vec \omega}| n)
\over |{\vec \omega}| } + {\epsilon_{\beta \alpha \gamma} \delta_{\alpha}
\omega_{\gamma} \over \omega^2} \cos(|{\vec \omega}|n) \Biggr\},
\label{position}
\end{equation}
with $\omega^2 = \omega_{\alpha}\omega_{\alpha}$.
Examination of (\ref{position}) shows that the geometric center moves on
average in the direction ${\bf e}_{\beta}(0)\omega_{\beta}/|{\vec \omega}|$ and
traces out a helix with helix axis offset $r = |{\vec \omega} \times {\vec
\delta}|/
|{\vec \omega}|$ and pitch $p = 2 \pi {\vec \delta} \cdot
{\vec \omega}/\omega^2$.
The shift vector, rotation vector, helix axis offset, and pitch are
listed in Table~\ref{shapes} for each of the three shapes of B-form DNA
we are considering.

\begin{table}
\caption{Geometric Properties of Several Helices}\vspace{0.2in}
\begin{center}
\begin{tabular}{ccccc}
Shape & ${\vec \delta}$ & ${\vec \omega}$ & Helix Axis Offset & ${\rm
Pitch}/2\pi$\\
\hline
I & $(0,0,\delta_3)$ & $(0,0,\omega_3)$ & $0$ & $\delta_3/\omega_3$\\
II & $(0,\delta_2, \delta_3)$ & $(0,0,\omega_3)$ & $\delta_2$
& $\delta_3/\omega_3$\\
III & $(0,0,\delta_3)$ & $(0,\omega_2,\omega_3)$
  & $\omega_2 \delta_3/\sqrt{\omega^2_2 + \omega^2_3}$ & $\delta_3
\omega_3/(\omega^2_2 + \omega^2_3)$
\end{tabular}
\end{center}
\label{shapes}
\end{table}

In a deformed state, the displacement and rotation rates change to
${\vec \delta}' = {\vec \delta} + {\vec \Delta}(n)$ and ${\vec
\omega}' = {\vec \omega} + {\vec \Omega}(n)$ and define new positions
${\vec x}'(\eta_1, \eta_2, n)$ for points on plate $n$.  Our goal is
to calculate the energy of this deformed state relative to the
equilibrium helical state to second order in the small parameters
${\vec \Delta}$ and ${\vec \Omega}$, which can in general depend on
$n$.  We consider plates of thickness $\delta n$ and calculate the
energy in the deformed state of each plate to order $\delta n$,
ignoring terms of order $(\delta n)^2$ or higher.  We then sum over
all plates to obtain the total energy of the rod.  Since the energy of
each plate is proportional to $\delta n$, the sum over all plates can
be converted into an integral over $n$.

The energy of the $n$th plate will depend on the values of ${\vec
\Delta}$ and ${\vec \Omega}$ at $n$ and, to the order we consider,
{\it not} on their derivatives with respect to $n$.  The energy of each plate
will be the same function of ${\vec \Delta}$ and ${\vec
\Omega}$ for every $n$, so we need only calculate the energy of a
single reference plate.  Our elastic theory provides us with
distortion energies of the reference plate as a function of the
strains $u_{ij} = {1 \over 2}(\partial_iu_j + \partial_j u_i)$ ($i,j =
x,y,z$), where ${\vec u} = {\vec x}' - {\vec x}$ is the displacement
variable and $x$,$y$, and $z$ are the Cartesian coordinates of the
reference plate.  By a suitable choice of orientation we may take
${\bf e}_1 = {\hat x}$, ${\bf e}_2 = {\hat y}$, and ${\bf e}_3 = {\hat
z}$ for the plate of interest.  The free energy for a single plate
with thickness $\delta z$ is (in units of $k_B T$):
\begin{equation}
{F_{\rm plate} \over k_B T} = {1 \over 2} \int^{z + \delta z}_z \int dx dy dz~
K_{ijkl}u_{ij}u_{kl}.
\label{singleplate}
\end{equation}

Our goal is to express this energy as a function of ${\vec \Delta}$
and ${\vec \Omega}$ and change integration variables from $x$, $y$, and
$z$ to $\eta_1$, $\eta_2$, and $n$.  The Jacobian relating the two coordinate
systems is found by taking $\eta_{\alpha}$
and $n$ derivatives of ${\vec x}(\eta_1, \eta_2, n)$ and remembering that
${\bf e}_{\alpha} \cdot {\bf e}_{\beta} = \delta_{\alpha \beta}$.
\begin{equation}
dx dy dz = |\delta_3 + \epsilon_{3\beta\gamma}
\omega_{\beta}\eta_{\gamma}| d\eta_1 d\eta_2 dn.
\label{jacobian}
\end{equation}
To calculate the strain tensor $u_{ij}$ we must take the
derivatives of ${\vec u}(\eta_1,\eta_2,n)$ with respect to
$\eta_{\alpha}$ and $n$ and then relate these coordinates to Cartesian
coordinates $x$, $y$, and $z$ fixed on the plate.  Since the
derivatives of $u_i$ with respect to $\eta_{\alpha}$ only include
terms proportional to $\delta n$, the only derivatives in the strain
tensor contributing to the total free energy in (\ref{singleplate})
are those in the $n$-direction.  We therefore only need to calculate
$d{\vec u}/dn$.  Using the equations of motion,
(\ref{triad}) and (\ref{tangent}), and the expression for ${\vec
x}(\eta_1, \eta_2, n)$, we have to linear order in $\Omega_{\alpha}$
and $\Delta_{\alpha}$
\begin{equation}
{d {\vec u}(\eta_1, \eta_2, n) \over dn} = {\bf e}_{\alpha}
\Bigl[ \Delta_{\alpha} + \epsilon_{\alpha \beta \gamma} \Omega_{\beta}
\eta_{\gamma} \Bigr] + (\delta_{\alpha} + \epsilon_{\alpha \beta \gamma}
\omega_{\beta}\eta_{\gamma})\Bigl[ {\bf e}^{'}_{\alpha} -
{\bf e}_{\alpha} \Bigr].
\label{strain}
\end{equation}
We can neglect the last terms in (\ref{strain}) since they also
are proportional to $\delta n$.  (We note that deformations of the
base pairs can be included if we retain these $\delta n$
contributions.)

We choose the slab at $n$ to have its internal triad point along
a Cartesian coordinate system $(x,y,z)$ fixed on the plate with the
$z$-direction aligned with ${\bf e}_3$ and the $x$- and $y$-directions
aligned with ${\bf e}_1$ and ${\bf e}_2$ respectively.  To leading
order in $\Omega_{\alpha}$ and $\Delta_{\alpha}$, the only relevant
components of the strain tensor come from the derivatives
\begin{equation}
\partial_z u_i = {\partial u_i \over \partial n}{\partial n \over
\partial z} = (\Delta_i + \epsilon_{ijk} \Omega_j \eta_k)/\delta_3,
\label{invariant}
\end{equation}
where $\partial n/\partial z = 1/\delta_3$ for the three shapes we are
considering.  In (\ref{invariant}) the sums over $\alpha$, $\beta$,
and $\gamma$ have been replaced by sums over $i$, $j$, and $k$ because
we chose the body-fixed coordinates to be along $(x, y, z)$.  We have
now reduced the number of effective elastic constants from $13$ to $4$
since only $z$ derivatives contribute to the free energy.  The free
energy of a single plate in terms of the strains
${\vec \Delta}$ and ${\vec \Omega}$ and the coordinates $\eta_1$,
$\eta_2$, and $n$ is:
\begin{equation}
{F_{\rm plate} \over k_B T}  = \int^{n + \delta n}_n dn~\int d^2\eta~
{ |\delta_3 + \epsilon_{3\beta\gamma}\omega_{\beta} \eta_{\gamma}|
\over 2\delta^2_3} \Bigl\{ K_{ij}  (\Delta_i + \epsilon_{ilm}
\Omega_l \eta_m) (\Delta_j + \epsilon_{jkn} \Omega_k \eta_n) \Bigr\},
\label{freeenergy}
\end{equation}
where $K_{ij}$ is related to $K_{ijkl}$ by
\begin{equation}
K_{ij} = \left\{
	\begin{array}{ll}
		{1 \over 4}K_{zizj} & i = j = x~{\rm or}~i = j = y\\
		{1 \over 2}K_{zizj} & i = y,~j = z \\
		K_{zizj} & i = j = z.
	\end{array}
	\right.
	\label{Krelation}
\end{equation}
We see that our elasticity theory of DNA as a stack of thin, rigid
plates has four contributing elastic constants, $K_{xx}$, $K_{yy}$,
$K_{yz}$, and $K_{zz}$.  Below we will refer to these elastic
constants as $K_{11}$, $K_{22}$, $K_{23}$, and $K_{33}$, respectively,
since the plate-fixed axes rotate in space as we move along the
helical stack.  We note that an isotropic theory has only two
contributing constants, $K = K_{11}/2 = K_{22}/2$ and $K_{33}$.  The
total free energy for the collection of plates is obtained by adding
up the contributions to the free energy from each plate.  The total
free energy will have the same form as
(\ref{freeenergy}) with the Cartesian indices on the strains $(x,y,z)$
replaced by the plate-fixed indices $(1,2,3)$.

\section{DNA Effective Free Energy}

We now sum up the contributions from all slices and do the integrals over
$\eta_1$ and $\eta_2$ in (\ref{freeenergy}) to get an effective theory
for the DNA molecule.  Before we actually do this, we will make
some further simplifications and approximations.  We first use the fact that
under the DNA symmetry operation $\eta_2$ switches sign and hence all
moments odd in $\eta_2$ are zero:  $\langle\eta_2\rangle =
\langle\eta^m_1\eta_2\rangle = 0$ ($m = 1,2,3,...$) where
$\langle\cdot\rangle \equiv {\cal A}^{-1}\int~d^2\eta~(\cdot)$ and
${\cal A}$ is the cross sectional area in the $\eta_1-\eta_2$ plane.
We also locate the origin of each plate at its geometric center so
that $\langle \eta_1 \rangle = 0$.

In the unstressed configuration of B-form DNA the base pairs make an
angle of close to $\pi/2$ with the molecular axis\cite{saenger}.  We
therefore take $\omega_{1,2}~l_0
\ll \delta_3$ where $\delta_3 \approx 3.4$~\AA~ is the axial rise of the
base pairs.  In addition, the geometric
centers of the plates are roughly located on the molecular axis
\cite{saenger}; we therefore also assume $\delta_{1,2}/\delta_3 \ll 1$.  We
will characterize the unstressed straight state of B-DNA by two
known macroscopic parameters (the axial rise $\delta_3$ and the twist rate
of the plates about the molecular axis $\omega_3 = 2 \pi \delta_3/l_0
\approx 0.63~{\rm rad}$) and two unknown microscopic parameters ($\delta_2$ and
$\omega_2$).  In what follows we will assume that $\delta_2/\delta_3$,
$\omega_2/\omega_3$, and the second moments $\langle \eta^2_1 \rangle/l^2_0$
and $\langle \eta^2_2 \rangle/l^2_0$ are small and work to quadratic
order in these quantities.

The effective free energy is given below in units of $k_BT$.  We
neglect the $\omega_2$ contribution arising from the Jacobian in
(\ref{freeenergy}) since it yields terms that are third order
in $\omega_2/\omega_3$, $\langle \eta^2_1
\rangle/l^2_0$, and $\langle \eta^2_2 \rangle/l^2_0$.  Including these
terms does not alter our findings below for the twist-stretch
coupling:
\begin{eqnarray}
{F_{\rm DNA} \over k_B T} = ({\cal A}/2\delta_3) \int dn~\biggl\{
K_{11} \Delta^2_1 + K_{22}\Delta^2_2 + K_{33} \Delta^2_3 + K_{33}
\langle\eta^2_2
\rangle\Omega^2_1 +  \nonumber \\
K_{33} \langle\eta^2_1\rangle\Omega^2_2 +~\biggl(
K_{11}\langle\eta^2_2\rangle + K_{22}\langle\eta^2_1\rangle\biggr)
\Omega^2_3 - 2K_{23}\langle\eta^2_1\rangle\Omega_2\Omega_3 +
2K_{23}\Delta_2\Delta_3 \biggr\}.
\label{general}
\end{eqnarray}
We note from (\ref{general}) that the elastic constant $K_{33}$ acts
as a stretch modulus and a bend modulus.

\section{Twist-Stretch Coupling}

Motivated by recent stretching experiments on torsionally
constrained single molecules of DNA\cite{Strick}, we rewrite
(\ref{general}) in terms of the relative overtwist $\sigma$ about the
molecular axis and the relative stretch $\epsilon$ along the molecular
axis.  To accomplish this we must first find $\epsilon$ and $\sigma$
in terms of the rotation and shift variables $\Omega_i$ and
$\Delta_i$.  After we incorporate $\epsilon$ and $\sigma$ into
(\ref{general}), we minimize over the remaining unconstrained
variables.  This gives us the twist-stretch energy in terms of the
elastic constants $K_{ij}$ and the geometric parameters of the DNA
helix.  We can then identify the stretch modulus $B$, the twist
modulus $C$, and the twist-stretch modulus $D$.  We note that this
twist-stretch energy neglects the effects of thermal fluctuations;
these effects have been studied recently\cite{Mezard,Moroz}.

We now derive the twist-stretch coupling for the three shapes shown
in Table~\ref{shapes} by considering the general case $\omega_2,
\delta_2 \ne 0$ and then neglecting $\omega_2\delta_2$ cross terms.
The unstressed configuration is therefore ${\vec \omega} =
(0,\omega_2,\omega_3)$ and ${\vec \delta} = (0,\delta_2,\delta_3)$.
Each of the three equilibrium shapes mentioned previously can be
obtained from this parameterization, for instance, shape II is
generated by taking $\delta_2 \ne 0$ and $\omega_2 = 0$.
We define the stretch along the molecular axis as the relative
deviation in the extension $L$ along the molecular axis from its
unstressed value $L_0$,
\begin{equation}
\epsilon = {L \over L_0} - 1.
\label{excessstretch}
\end{equation}
To find $\epsilon$ in terms of $\Delta_i$ and $\Omega_i$
we must determine the extension $L$ from (\ref{position}).
We see that each base pair step increases the extension by an amount
${\vec \delta} \cdot {\vec \omega}/|{\vec \omega}|$ in the direction
${\vec \omega}/|{\vec \omega}|$.  The total extension for
$N$ steps in the distorted configuration is therefore
\begin{equation}
L =  N {\vec \delta}' \cdot {{\vec \omega}' \over |{\vec \omega}'|}
\label{extension}
\end{equation}
where the primes refer to the distorted state, {\it i.e.} ${\vec \delta}'
= (\Delta_1, \delta_2 + \Delta_2, \delta_3 + \Delta_3)$.
We can now expand (\ref{excessstretch}) to linear order in the
$\Omega_i$ and $\Delta_i$ and to quadratic order in the
small parameters $\delta_2/\delta_3$ and $\omega_2/\omega_3$:
\begin{equation}
\epsilon = {\Delta_3 \over \delta_3} + \Bigl({\delta_2 \over \delta_3} -
{\omega_2 \over \omega_3}\Bigr) {\Omega_2 \over \omega_3} +
{\omega_2 \over \omega_3}{\Delta_2 \over \delta_3} +
{\omega^2_2 \over \omega^2_3}{\Omega_3 \over \omega_3}.
\label{stretch}
\end{equation}
Note that when the helix axis offset of the helical stack is zero $(\delta_2 =
\omega_2 = 0)$, the stretch reduces to $\epsilon = \Delta_3/\delta_3$.
We also note that when $\omega_2 = 0$ $\Delta_3$ and
$\Omega_2$ are the only distortions contributing to the stretch.

We must express the relative excess link $\sigma = (Lk/Lk_0) - 1$ in
terms of the $\Omega_i$, where the linking number $Lk$ is the number
of times the plates rotate about the molecular axis over the length of
the stack.  In its unstressed state, DNA has linking number $Lk_0 =
L_0/l_0$.  (We are working in the force regime where the molecular
axis is nearly straight, and hence link and relative twist about the
molecular axis are synonymous.)  We see from (\ref{position}) that it
takes $2\pi/|{\vec
\omega}|$ steps to complete one revolution about the molecular axis.
Therefore in $N$ steps the top plate has rotated by $N |{\vec
\omega}|/2\pi$ relative to the bottom plate.  The relative excess twist
about the molecular axis is therefore
\begin{equation}
\sigma = {|{\vec \omega}'| \over |{\vec \omega}|} - 1,
\label{sigma}
\end{equation}
where ${\vec \omega}' = (\Omega_1, \omega_2 + \Omega_2, \omega_3
+ \Omega_3)$.  Expanding (\ref{sigma}) to linear order in the $\Omega_i$
and to quadratic order in $\omega_2/\omega_3$, we find
\begin{equation}
\sigma = {\omega_2 \over \omega_3} {\Omega_2 \over \omega_3} +
\Bigl( 1 - {\omega^2_2 \over \omega^2_3} \Bigr) {\Omega_3 \over \omega_3}.
\label{sigexpression}
\end{equation}
Note that when $\omega_2 = 0$, the base pairs are perpendicular to
the molecular axis and the relative excess twist is simply $\Omega_3/\omega_3$.

Solving (\ref{stretch}) for $\Delta_3$ and (\ref{sigexpression})
for $\Omega_3$, we substitute into (\ref{general}) and
minimize over $\Delta_{1,2}$ and $\Omega_{1,2}$.  This gives the
effective twist-stretch energy per length along the molecular
axis:
\begin{equation}
{F_{TS}(\epsilon,\sigma) \over N \delta_3 k_B T} =
{{\cal A} \over 2}\biggl( B \epsilon^2 + C \sigma^2 + 2 D \epsilon
\sigma \biggr),
\label{twiststretch}
\end{equation}
where $N$ is the total number of plates, $B$ is the stretch modulus,
$C$ is the twist modulus, and $D$ is the twist-stretch modulus.  To
zeroth order in the small parameters $\delta_2/\delta_3$ and
$\omega_2/\omega_3$ the stretch and twist moduli are:
\begin{eqnarray}
B & = & K_{33} - K^2_{23}/K_{22} \nonumber \\
C & = & (\omega^2_3/\delta^2_3)\Bigl(K_{11}\langle\eta^2_2\rangle +
\langle\eta^2_1\rangle(K_{22} - K^2_{23}/K_{33})\Bigr).
\label{bulkmoduli}
\end{eqnarray}
$B$, $C$, and $D$ have dimension $L^{-3}$ and therefore the
coefficients of the quadratic dimensionless strains in
(\ref{twiststretch}) scale as a persistence length divided by the square
of the helical pitch as found previously\cite{Kamien}.  The
twist-stretch couplings $D_i$ for the three shapes given in
Table~\ref{shapes} are shown below to lowest order in the helix axis
offsets $r_i$:
\begin{eqnarray}
D_{\rm I} & = & 0 \nonumber \\
D_{\rm II} & \approx & {K_{23} \over K_{33}}\Bigl(K_{33} -
{K^2_{23} \over K_{22}}\Bigr) {r_{\rm II} \over \delta_3} \nonumber \\
D_{\rm III} & \approx & {K_{23} \over K_{33}}\Bigl(K_{33} - {K^2_{23} \over
K_{22}}\Bigr) {r_{\rm III} \over \delta_3} + \Bigl(K_{11}
{\langle\eta^2_2\rangle \over \langle\eta^2_1\rangle} + K_{22} - K_{33}\Bigr)
\Bigl({r_{\rm III} \over \delta_3}\Bigr)^2,
\label{TSmodulus}
\end{eqnarray}
where the helix axis offsets for each shape are $r_{\rm I} = 0$, $r_{\rm II} =
\delta_2$, and $r_{\rm III} \approx \omega_2 \delta_3/\omega_3$.  In performing
the calculation of the twist-stretch moduli we have made two
simplifications: 1) we have dropped the quadratic terms in the helix
axis offset that are proportional to $K_{23}$ and 2) we have not included the
$\omega_2$ contributions arising from the Jacobian in
(\ref{freeenergy}).  This last simplification does not affect our
conclusions concerning the twist-stretch coupling (TS) shown in
Table~\ref{TS} since $\omega_2$ contributions can only affect the
twist-stretch coupling of shape III and shape III already has
twist-stretch couplings in both the isotropic and anisotropic models.

\begin{table}
\caption{Determination of the twist-stretch coupling
for each shape of B-form DNA is shown below.  The symbol TS signifies
that there is a nonzero twist-stretch coupling.}
\vspace{0.2in}
\begin{center}
\begin{tabular}{cccc}
 & Helix Axis & Isotropic Model & Anisotropic Model \\
Shape & Offset & $(K_{23} = 0)$ & $(K_{23} \ne 0)$\\
\hline
${\rm I}$ & $0$ & $0$ & $0$\\
${\rm II}$ & $\delta_2$ & $0$ & TS\\
${\rm III}$ & $\omega_2 \delta_3/\omega_3$ & TS & TS
\end{tabular}
\end{center}
\label{TS}
\end{table}

We see from (\ref{TSmodulus}) that the twist-stretch coupling in our
rigid plate model vanishes when the helix axis offset $r$ vanishes.
However, calculations for a pretwisted rod of material with a
three-dimensional isotropic elasticity yield a twist-stretch coupling
for zero helix axis offset provided the rod cross-section is not
circular\cite{Krenk}.  To achieve consistency with our results, it
seems likely that the magnitude of this coupling decreases as the
elastic constants coupling in-plane strains ({\it i.e.}, $u_{xx}$,
$u_{xy}$, $u_{yy}$) increase.  We can thus view our calculation as the
limit when these elastic constants diverge.

We also see that in the isotropic limit ($K_{23} = 0$) the
twist-stretch coupling for shape III scales quadratically with the
small helix axis offset as found previously in
Refs.~\cite{Kamien,Ohern}.  However, when the $K_{23}$ anisotropy is
nonzero, the twist-stretch coupling scales linearly with the small
helix axis offset.  We note that the twist-stretch coupling in the
isotropic limit vanishes for shape II since it has $\omega_2 =0$.
However, shape III yields a twist-stretch coupling in both the
isotropic and anisotropic cases since it has $\omega_2 \ne 0$.  This
may be explained by considering the manner in which the plates are
stacked in the two different geometries.  In shape II the plates are
stacked perpendicular to the molecular axis.  In this case, the plates
can move along the molecular axis without rotating in response to the
stretching force.  In shape III the plates are not perpendicular to
the molecular axis, but are instead perpendicular to the local tangent
$d{\vec r}/dn$.  In order to align the plates perpendicular to the
molecular axis, they must twist.  This implies that an isotropic rod
${\it bent}$ into the shape of a helix will have a twist-stretch
coupling, but carving a helical shape out of isotropic material will
not produce a twist-stretch coupling.

\section{Comparison to the Ribbon Model of DNA}

We compare our plate model to a ribbon model of DNA
recently discussed in Ref.~\cite{Ohern}.  There we modeled DNA
as a thin helical ribbon that is only allowed to stretch along the
ribbon axis.  Changes in extension along the molecular axis arose from
changes in the angle the ribbon tangent makes with the molecular axis
and from stretching the ribbon axis.  We showed that the
twist-stretch coupling scaled quadratically with the helix axis offset if
the twist-bend, twist-stretch, and bend-stretch couplings for twists and bends
about and stretches along the {\it ribbon axis} were set to zero.

To make a direct comparison between the two models we consider the
unstressed configuration ${\vec \delta} = (0,0,\delta_3)$ and ${\vec
\omega} = (0,\omega_2,\omega_3)$ and do not allow stretch along the
$1$ and $2$ axes of each plate by setting $\Delta_1 = \Delta_2 = 0$.
This corresponds to a helical stack of plates with the planes of the
plates perpendicular to the tangent to the curve ${\vec r}$ connecting
the geometric centers of the plates and stretching only allowed along
the tangent $d{\vec r}/dn$.  When we set $\Delta_1 = \Delta_2 = 0$ in
(\ref{general}), we find a free energy identical to the one studied
previously\cite{Ohern}.  We see below that our
expression for the free energy admits all of the couplings found in
the ribbon model, namely the bend-stretch $(\Omega_2 \Delta_3)$,
twist-bend $(\Omega_2
\Omega_3)$, and twist-stretch $(\Omega_3 \Delta_3)$ couplings.
(We have included the $\omega_2$ contributions from the Jacobian and
kept each coupling to lowest order in $\omega_2/\omega_3$ and the
second moments.)
\begin{eqnarray}
\lefteqn{ {F_{\rm ribbon} \over k_B T}~= } \nonumber \\
& & ({\cal A}/2\delta_3) \int  dn \biggl\{K_{33} \langle\eta^2_2
\rangle\Omega^2_1 + K_{33}
\langle\eta^2_1\rangle\Omega^2_2 +~\biggl( K_{11}\langle\eta^2_2\rangle +
K_{22}\langle\eta^2_1\rangle\biggr) \Omega^2_3 + K_{33}
\Delta^2_3 + \nonumber \\
& &  2 K_{33} {\omega_2 \over \delta_3}\langle\eta^2_1\rangle
\Omega_2 \Delta_3 - 2K_{23}\langle\eta^2_1\rangle\Omega_2\Omega_3
- 2 K_{23}{\omega_2 \over \delta_3}\langle\eta^2_1\rangle\Omega_3\Delta_3
\biggr\}.
\label{ribbon}
\end{eqnarray}
We find that the presence of the the last two terms leads to
twist-stretch couplings that scale linearly with the helix axis offset
$D \sim (K_{23}/K_{33})r_{\rm III}/\delta_3$.  If we set $K_{23} = 0$,
the lowest order contribution is quadratic in the helix axis offset
and we obtain the result $D \sim K_{33}(r_{\rm III}/\delta_3)^2$ found
previously\cite{Kamien,Ohern}.

\section{Conclusion}

We have presented an elasticity theory for B-form DNA modeled as a
stack of thin, rigid plates.  The symmetry of the DNA molecule and the
assumption of plate rigidity enabled us to reduce the number of
contributing elastic constants to four.  We then derived an effective
twist-stretch energy for DNA in terms of the relative excess twist
$\sigma$ about the molecular axis and the stretch $\epsilon$ along the
molecular axis.  We found an effective twist-stretch coupling with
terms that scale linearly and quadratically with the small helix axis
offset.  We have argued that deformations of the base pairs are more
costly than deviations in the shifts and rotations of the base pairs
from their unstressed values.  We therefore conclude that in the force
regime where deformations of the plates are negligible, the
twist-stretch coupling is due to a nonzero helix axis offset.  We have
also found that shape II, in which the normals to the base pairs parallel
to the molecular axis, does not have a twist-stretch coupling in the
limit $K_{23} = 0$.  Since recent experiments show that there is a
large twist response to stretch, this implies that either the coupling
$K_{23}$ is relevant for an elastic description of B-DNA or that the
unstressed straight state of B-DNA is composed of base pairs that tilt
relative to the molecular axis.

An important next step is to estimate the elastic constant $K_{23}$ to
determine the relative magnitudes of the isotropic and anisotropic
terms in the twist-stretch coupling.  One might also calculate the
twist-stretch coupling for DNA plasmids where the unstressed state of
the molecular backbone is circular rather than linear.  This
calculation could provide a theoretical estimate of the twist-stretch
coupling found in experiments on complexes of Rec-A protein with DNA
plasmid\cite{stasiak}.

\begin{center}
{\bf Acknowledgments}
\end{center}
We would like to thank D. Bensimon, J.F.~Marko, and J.D.~Moroz for
helpful comments.  CSO, RDK, and TCL were supported in part by NSF
grant DMR94--23114.  RDK was also supported in part by NSF grant
DMR95--20175 and PN was partially supported by NSF grant DMR95--07366.

\vfill\eject
\begin{center}
{\bf Appendix: Translational Invariance of the Plate Origin}
\end{center}

We comment here on the invariance of our model under the
choice of the plate origin.  This is an important feature since a
twisted stack of plates can be described by an infinite number of
one-dimensional curves depending on the choice of the reference
point on each plate.  We
proceed by considering a change in the origin, or, in particular a
change in the coordinate system such that ${\vec x} = (\eta_1,\eta_2,s)
\rightarrow \tilde x = (\eta_1, \eta_2, s) + {\vec R}$ where ${\vec R}$
is an arbitrary vector.  While this translation does not affect the
relative rotation of two consecutive plates ${\vec \omega}$, it does
affect the {\it relative translation}.  Consider the action of
$g = ({\vec \omega},{\vec \delta})$ on ${\vec x}$:
\begin{equation}
{\vec x} \mathop{\longrightarrow}\limits^{g} {\vec x} + {\vec \delta}
- {\vec \omega} \times {\vec x}.
\label{transform}
\end{equation}
We now consider the transformation in (\ref{transform}) shifted by
${\vec R}$ and compare it to the transformation $\tilde g = (\tilde
\omega,\tilde \delta)$ written in terms of the new coordinate system
$\tilde x$.
\begin{equation}
{\vec x} + {\vec R} = \tilde x + \tilde \delta - \tilde \omega \times
\tilde x \equiv \biggl[ \biggl( \tilde x - {\vec R} \biggr) +
{\vec \delta} - {\vec \omega} \times \biggl( \tilde x - {\vec R} \biggr)
\biggr] + {\vec R}
\label{rewrite}
\end{equation}
In order for $g$ and $\tilde g$ to act the same way, we must
choose $\tilde \omega = {\vec \omega}$ and $\tilde \delta = {\vec
\delta} - {\vec \omega} \times {\vec R}$.
We note that not only does $({\vec \omega}, {\vec \delta})$ transform
as above, but in addition, due to the linearity of the transformation,
$({\vec \Omega}, {\vec
\Delta})$ transforms the same way.  As a result, we see that the
derivatives of interest in
(\ref{invariant}) transform as
\begin{equation}
\partial_z u_i = \bigl[\Delta_i - \epsilon_{ijk}\Omega_j\eta_k\bigr]/\delta_3
\longrightarrow \biggl[ \Delta_i + \epsilon_{ijk}\Omega_j\biggl( \eta_k -
R_k\biggr) + \epsilon_{ijk}\Omega_j{\vec R}^{tr}_k\biggr]/\delta_3,
\label{invartrans}
\end{equation}
where ${\vec R}^{tr}_k$ is the projection of ${\vec R}$ onto the
$\eta_1-\eta_2$ plane.  Since changing the base point on each slab
only requires components of ${\vec R}$ in that plane, we see that the
last two terms in (\ref{invartrans}) cancel and the relevant
derivatives are invariant with respect to the choice of the origin on
the plate.  Thus we are free to choose the most convenient origin for
any calculation of interest.

\newpage

\begin{thebibliography}{10}

\bibitem{Siggia}
Marko J.F. and  Siggia E.D., Macromolecules {\bf 24} (1994) 981-988.

\bibitem{stretchdna}
Marko J.F. and Siggia E.D., Macromolecules {\bf 28} (1995) 8759-8770.

\bibitem{fain}
Fain B., and Rudnick J., ``Conformations of Linear DNA'', preprint
(cond-mat/9610126), (1996).

\bibitem{Markonew}
Marko J.F., Europhys. Lett. {\bf 38} (1997) 183.

\bibitem{Kamien}
Kamien R.D., Lubensky T.C., Nelson P, and O'Hern C. S.,
Europhys. Lett. {\bf 38} (1997) 237-242.

\bibitem{Mezard}
Bouchiat C. and  M\'{e}zard M., preprint (cond-mat/9706050), (1997).

\bibitem{smith2}
Smith S.B., Cui Y.J., and Bustamante C., Science {\bf 271} (1996) 795-799.

\bibitem{Wang}
Wang M.D., Yin H., Landick R., Gelles J., and Block S.M.,
Biophysical Journal {\bf 72} (1997) 1335-1346.

\bibitem{stasiak}
Stasiak A. and Di Capua E., Nature {\bf 299} (1982) 185-186.

\bibitem{Strick}
Strick T.R., Allemand J.F., Bensimon D., Bensimon A., and Croquette V.,
Science {\bf 271} (1996) 1835-1837.

\bibitem{cluzel}
Cluzel P. {\it et. al.}, Science {\bf 271} (1996) 792-794.

\bibitem{Ohern}
Kamien R.D., Lubensky T.C., Nelson P, and O'Hern C. S.,
``Twist-Stretch Elasticity of DNA'', Proceedings of MRS Fall 1996
Meeting, Boston, MA.

\bibitem{Moroz}
Moroz J.D. and Nelson P., preprint, (1997).

\bibitem{philnote}
The helical geometry of the DNA molecule alone cannot explain the
twist-stretch coupling. Consider straightening out the helical
sugar-phosphate backbones without changing the length per phosphate
$0.6~$nm or the distance from the center of the molecule $0.9$~nm.
Since each base pair step is $h=0.34\,$nm, the circumferential
length per step is $l_c=\sqrt{.6^2-.34^2}\,$nm. The corresponding
twist angle per step is given by \cite{Calladine} $\theta= 2
\arcsin((l_c/2)/.9{\rm nm})=32^\circ$.  An extension of the molecule
by $\delta h/h=\epsilon$, yields an untwisting by an amount
$\sigma=\delta\theta/\theta=\sigma_{\epsilon=0}-\epsilon/2.0$.  The
twist response to stretch given by the coefficient
of $\epsilon$ in $\sigma$ versus $\epsilon$ is much smaller than the
coefficient ($\approx 7$) found experimentally\cite{Strick}.

\bibitem{Calladine}
Calladine C. R. and Drew H. R., \emph{Understanding DNA: the molecule
and how it works}, Academic, London (1992).

\bibitem{Lavery}
Lavery R. and Sklenar H., Journal of Biomelecular Structure \&
Dynamics {\bf 6} (1989) 655-667.

\bibitem{saenger}
Saenger W., \emph{Principles of Nucleic Acid Structure}, Springer-Verlag,
New York (1984).

\bibitem{Krenk}
Krenk S., Journal of Applied Mechanics {\bf 50} (1983) 137-142.

\end{thebibliography}
\end{document}